\documentclass[aps,pra,showpacs,twocolumn]{revtex4-1}
\usepackage[dvips]{graphicx}
\usepackage{amsmath,amssymb}
\usepackage{amsfonts}
\usepackage{color}
\usepackage{amsmath}

\usepackage{graphicx}
\newcommand{\beq}{\begin{equation}}
\newcommand{\eeq}{\end{equation}}

\newcommand{\beqar}{\begin{eqnarray}}
\newcommand{\eeqar}{\end{eqnarray}}

\newcommand{\Eq}[1]{Eq.~(\ref{#1})}
\newcommand{\Eqs}[1]{Eqs.~(\ref{#1})}

\newcommand{\rt}{\rm rot}

\newcommand{\al}{\alpha}
\newcommand{\be}{\beta}
\newcommand{\ga}{\gamma}
\newcommand{\de}{\delta}
\begin{document}

\title{
Quantum acousto-optic transducer for superconducting qubits 
}
\author{ Vitaly S. Shumeiko}
\affiliation{                    
Chalmers University of Technology - S-412 96 Gothenburg, Sweden 
}

\begin{abstract}
We propose theory for reversible quantum transducer connecting superconducting qubits and optical photons using acoustic waves in piezoelectrics. The proposed device consists of integrated acousto-optic resonator that utilizes stimulated Brillouin scattering  for phonon-photon conversion, and piezoelectric effect for coupling of phonons to qubits. We evaluate the phonon-photon coupling rate, and show that the required power of optical pump as well as the other device parameters providing full and faithful  quantum conversion are feasible for implementation with the state of the art integrated acousto-optics.
 \end{abstract}

\maketitle

\section{Introduction}

 Network of superconducting qubits interconnected and measured with microwave photons
constitutes a c-QED (circuit quantum electrodynamics) architecture of quantum processor \cite{Girvin}.
This architecture demonstrated performance of basic quantum algorithms  with up to 9 qubits on chip \cite{qubits1,qubits2,qubits3,qubits4}, the number of qubits is expected to further increase.
 However the microwave photons are not suitable for long distance quantum communication, the optical channels are required in order to connect remote qubit clusters \cite{Kimble,Lukin}.   Development of microwave to optic quantum interfaces is a vital step towards large scale quantum networks. Also, the transferring to optical domain of complex nonclassical photonic states produced by c-QED technology might be attractive for quantum metrology \cite{Metrology}.
 
The problem of  interfacing c-QED network and optical photons is challenging.
It has been extensively discussed theoretically \cite{Wallquist}, and a number of solutions involving nonlinear optomechanical interfaces  have been proposed \cite{Lukin,Safavi NJP 2011,Vitali 2011, Vitali 2012}. and experimentally tested \cite{Cleland, Lehnert2014}.    

In a parallel development, a strong piezoelectric
coupling of superconducting qubit to propagating surface acoustic waves (SAW) was experimentally demonstrated \cite{Gustafsson1, Gustafsson2}, the coupling rate being comparable to the one in the c-QED devices. In another experiment \cite{Leek} a SAW resonator demonstrated a high quality factor, $Q\sim 10^5$, comparable to the microwave resonators. 
These experiments suggest a possibility of  using GHz-frequency phonons  
for on-chip quantum communication  (circuit quantum acousto-dynamics, c-QAD) \cite{Rabl NJP2012, Vandersypen}. Attractive feature of c-QAD architecture is a possibility of efficient coupling of phonons to optical photons thus providing means for long distance quantum   communication.

In this paper  we propose and  theoretically investigate a quantum  acousto-optic transducer that 
utilizes the mechanism of stimulated Brillouin scattering (SBS) \cite{Chiao} as a tool for reversible phonon-photon  conversion. 

SBS is a fundamental physical effect of inelastic scattering of light by acoustic waves in the presence of strong resonant optical field \cite{BoydBook}.  
SBS is observed in a variety of liquid and solid media, and  widely used in acousto-optic devices \cite{SBSfibers,SBSNatComm,CarmonBook}.
Under SBS the wave vectors of the resonant optical and acoustic modes have comparable values, which allows  for efficient conversion of telecom optical photons and acoustic phonons in GHz frequency range compatible with the c-QAD technology. 

The questions we raise, and answer in this paper  are: (i) how strong is the phonon-photon coupling provided by SBS, and (ii) can full phonon-photon conversion be achieved with realistic intensity of optical pump?

Stimulation of acousto-optic interaction by a strong resonant field under SBS can be understood as a peculiar, hybrid form of a non-degenerate parametric resonance, which couples physically different fields. Similar to purely electromagnetic non-degenerate parametric resonance in c-QED cavities \cite{Flurin,Aumentado,Abdo,Simoen}, SBS appears as either amplification of optical and acoustic modes, or mode hybridisation and Rabi oscillation  \cite{Kippenberg2014,Simmonds2015}.  At cryogenic temperatures relevant for operation of superconducting qubits, the Rabi oscillation regime maintains full coherence  because of small acoustic attenuation.  This regime is proposed for the quantum acousto-optic conversion.

The physical interaction underlying Brillouin scattering is a nonlinear photo-elastic effect - a change of the dielectric constant of the resonator material under elastic deformation. This is a common effect for all materials, resulting in many cases in dominant optical nonlinearity \cite{BoydBook}. 
In resonators, the variation of dielectric constant produces effect similar to the optomechanical effect of displacement of cavity boundaries \cite{ClerkBook,Law}. Furthermore, a reciprocal to the photo-elastic effect  is a force exerted by gradient of the light energy on elastic medium. Thus the photo-elastic effect can be understood as a distributed, bulk analog of the physically similar boundary effect in optomechanics. 

The purpose of this paper is to formulate a quantum theory of SBS in integrated acousto-optic resonator, and identify conditions for the full reversible phonon-photon conversion.

\section{Acousto-optic resonator}

 The envisioned  device is illustrated in Fig.~1. The device in Fig.~1a is an extension of the setup of the transmon-SAW experiment \cite{Gustafsson2}. It consists of a high quality SAW resonator defined  on a surface of a piezoelectric crystal by Bragg mirrors.  Acoustic signal is excited by SAW interdigital transducer (IDT) that simultaneously serves as a capacitance for the transmon qubit and is connected to a microwave line. The SAW resonator is integrated with a high quality optical  resonator consisting of a 1-dimensional wave guid loop defined on the surface of the piezoelectric. The optical field is injected into and extracted from the resonator by  evanescent coupling to a fiber. The optical and acoustic waves coexist in this integrated resonator and interact due to the SBS mechanism.  
In Fig. 1b an integrated acousto-optic resonator with the ring geometry is shown, inspired by  
 the design of existing optomechanical devices \cite{Tang2012}.  Transducer can also be realized with optomechanical crystals \cite{PainterBook} fabricated with piezoelctric materials \cite{Cleland,Srinivasan}.

\begin{figure}[h]
\begin{center}
\includegraphics[width=0.4\textwidth]{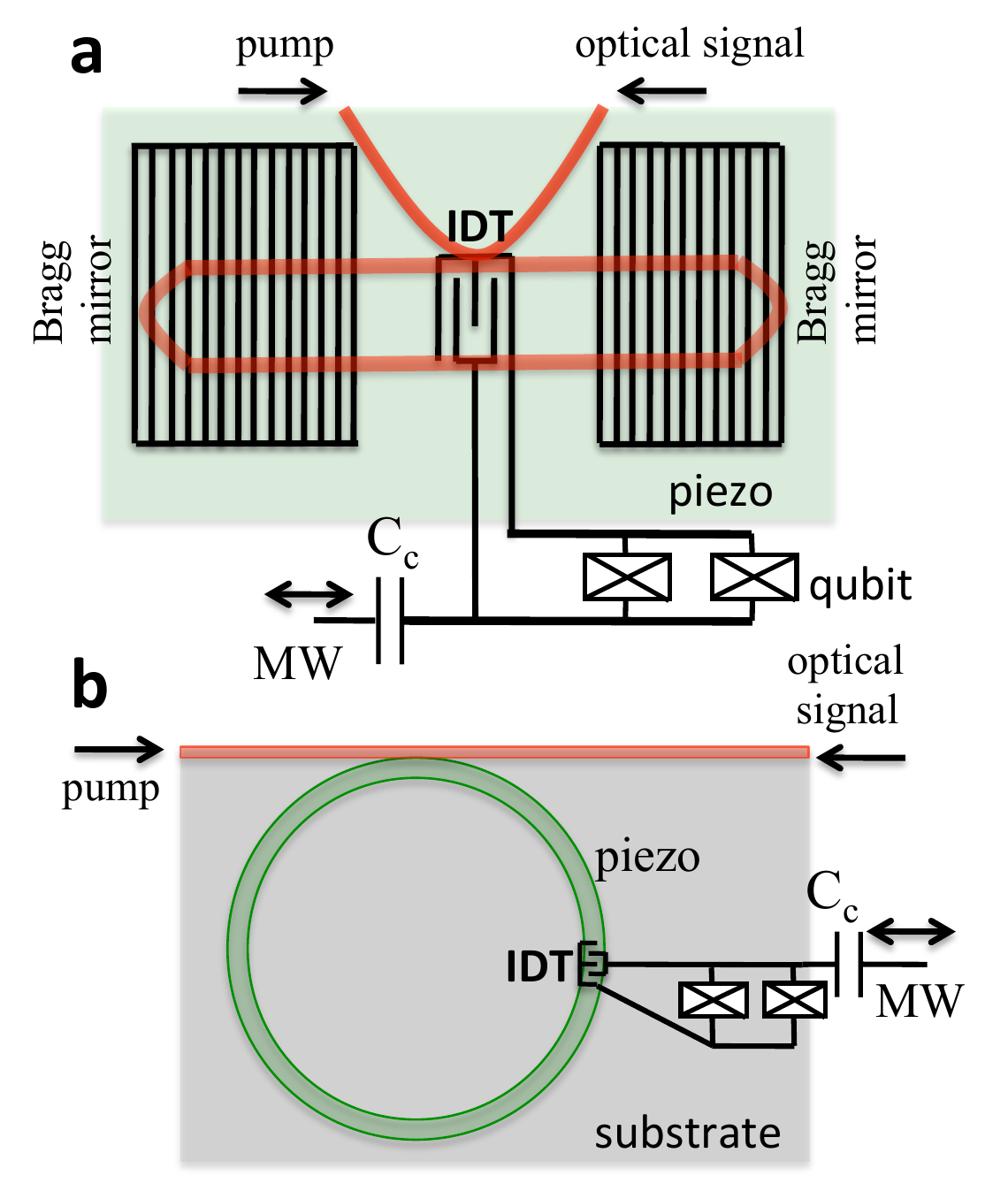}
\caption{(Color on-line) Acousto-optic transducer. {\bf a}: SAW resonator confined by the Bragg mirrors and integrated with the optical resonator (red line) defined at the surface of piezoelectric and evanescently coupled to a fiber; transmon qubit capacitor acts as IDT that couples microwave field to SAW, $C_c$ is a coupling capacitor to external microwave line (MW) \cite{Gustafsson2,Leek}.
{\bf b}: integrated SAW-optical ring resonator made with a piezoelectric film \cite{Tang2012}}.
\label{}
\end{center}
\end{figure}

\subsection{Classical theory of SBS}
  
Within the classical theory of SBS \cite{BoydBook,Kroll,Blombergen}, the photo-elastic interaction is introduced through variation of the dielectric  constant $\epsilon_{\al\be}$ under elastic deformation 
$ u_{\de,\ga}\,$, 
$\delta\epsilon_{\al\be} = \ga_{\al\be\ga\de} u_{\de,\ga}$, the interaction strength is quantified  with the photo-elastic coefficient tensor,  $\ga_{\al\be\ga\de}$ (indices denote spatial coordinates ($x,y,z$), index after coma indicates differentiation over respective variable, and convention is used for summation over repeated indices). \\
The equations of classical theory of SBS extended to piezoelectric materials read, 
\begin{eqnarray}\label{EOME}
&&\ddot D_{\al} +  c^2\,(\rt\rt \,E)_\al = 0\,, \quad  D_{\al,\al} =0\, \\
\label{EOMu1} 
 &&\rho \ddot u_{\ga} = c_{\al\be\ga\de} u_{\al,\be\de} +  e_{\al\be\ga}  E_{\al,\be} -   
 {\ga_{\al\be\ga\de}\over 8\pi} (E_{\al} E_{\be})_{,\de}  
 \\
\label{D}
&& D_\al = \epsilon_{\al\be} E_{\be} - 4\pi e_{\al\be\ga}  u_{\be,\ga} +  
\ga_{\al\be\ga\de} \, E_\be  u_{\ga,\de}\,,  \;\;\;\;
\end{eqnarray}
where $e_{\al\be\ga}$ is a piezoelectric tensor \cite{Auld}, $\rho$ is a mass density, $c_{\al\be\ga\de}$ is a stiffness tensor. 
These equations follow from a general Lagrangian for acousto-optic system \cite{Landau}, 
\begin{eqnarray}
{\cal L} &=& {1\over 2}\int dV \left[ {E^2-H^2\over 4\pi} + \varphi \,P_{\al,\al} + {1\over c}A_\al \dot P_\al \right.\nonumber\\
&+& \left. \dot u_\al^2 - c_{\al\be\ga\de}  u_{\al,\be}  u_{\ga,\de}\right] \,,
\end{eqnarray}
by computing  variations over electromagnetic potentials, $A_\al$ and $\varphi$, and displacement 
$u_\al$. Here 
$P_\al = \chi_{\al\be} E_\be - 2e_{\al\be\ga} u_{\be,\ga} +  (1/ 4\pi)\ga_{\al\be\ga\de} E_\be  u_{\ga,\de}$ is a macroscopic polarization vector, $\chi_{\al\be}$ is an electric susceptibility. 
The details of the derivation are presented in Appendix A.  

The last, nonlinear term in the elasticity equation (\ref{EOMu1}) describes the pressure exerted by the electromagnetic field on the elastic medium. This nonlinear term generates a mixture of optical modes, which may propagate with the speed of sound allowing for strong resonant interaction between acoustic and optical fields. This interaction is supported by resonant scattering of light in \Eq{EOME} by spatio-temporal grating formed by the sound. This is the classical wave picture of the Brillouin scattering.

Dynamics of the acousto-optic system, \Eqs{EOME}-(\ref{D}), consists of two different time scales associated with different propagation velocities of the light and the sound. These time scales are resolved by separating the fast transverse (optical) component of electric field,  $ E_\al^t = -(1/c)\dot A_\al$, $E^t_{\al,\al} =0$, 
\begin{eqnarray}\label{EOMA}
\epsilon_{\al\be} \ddot A_\be - c^2\,\Delta A_\al   + \ga_{\al\be\ga\de} \, 
\partial_t (\dot A_\be  u_{\ga,\de}) =0
\,,
\end{eqnarray}
and the slow longitudinal piezoelectric component,  
$E_\al^l = - \varphi_{,\al}$, $\rt \,E^l_\al =0$, coupled to the elastic field, 
\begin{eqnarray}\label{EOMu}
&& \!\! \rho \ddot u_{\ga} - c_{\al\be\ga\de} u_{\al,\be\de} -  e_{\al\be\ga} 
\varphi_{,\al\be}+ 
{\ga_{\al\be\ga\de}\over 8\pi c^2} \overline{(\dot A_{\al} \dot A_{\be}})_{,\de} = 0    \nonumber\\ 
 && \epsilon_{\al\be} \,\varphi_{,\al\be} + 4\pi e_{\al\be\ga} u_{\be,\ga\al}   =0  
 \, ,
\end{eqnarray}
the bar indicates averaging over fast optical oscillation. In \Eq{EOMu} a nonlinear  term containing small piezoelectric potential was omitted (for the details of derivation see Appendix A).  
The shortened equations (\ref{EOMA}) - (\ref{EOMu}) are associated with the Lagrangian, which we divide into the free field part,  ${\cal L}_0 $, and the photo-elastic interaction part, ${\cal L}_{int}$, 
\begin{eqnarray}\label{Lsimple}
{\cal L} &=&{\cal L}_0 + {\cal L}_{int} \,, \\
{\cal L}_0 &=& {1\over 2}\int dV\,\left\{ {1\over 4\pi c^2}  \epsilon_{\al\be} 
\dot A_\al \dot A_\be - {1\over 4\pi } A_{\al,\be}A_{\al,\be} + \rho \dot u_\al^2 \right. \nonumber\\
&-& \left.  c_{\al\be\ga\de} \, u_{\al,\be} u_{\ga,\de} 
+ 2e_{\al  \be\ga} \, u_{\be,\ga}\varphi_{,\al} + {1\over 4\pi} \epsilon_{\al\be} \, \varphi_{,\al}\varphi_{,\be} \right\} \nonumber\\
{\cal L}_{int} &=&  {1\over 2} \int dV\, {1\over 4\pi c^2} \ga_{\al\be\ga\de} \, u_{\ga,\de} 
\dot A_\al \dot A_\be \,.\nonumber
\end{eqnarray}
%

\subsection{Ring resonator}

The SBS equations are formulated for boundless continuous medium. To consider fields confined in a resonator, we focus for certainty on the ring resonator geometry of Fig. 1b, with the ring circumference much larger than the wave length, $L \gg \lambda$. The eigen modes of such a resonator can be approximated with the ones of a 1D wave guide with periodic boundary condition along 
$x$ direction. We expand the fields over the eigen modes of non-interacting resonator,
\begin{eqnarray}
A_\al(r,t) =  (1/ \sqrt{L}) \sum_{k,n} A_n(k,t)  e^{ikx}\psi_\al^n (r_\perp, k) \,, \\
u_\al(r,t) =
(1/ \sqrt{L}) \sum_{q,m} u_m(q,t)  e^{iqx} \,\phi_\al^m (r_\perp, q) \,,
\end{eqnarray}
were the normalized eigen functions, $\psi_\al^n (r_\perp, k)$ and  $\phi_\al^m (r_\perp, q)$, describe the transverse confinement of the respective fields, indices $n,m$ count the transverse eigen modes. 

The exact form of the transverse eigen functions is dictated by particular geometry of the wave guide. Computing these  functions analytically is particularly difficult for the acoustic modes, the difficulty results from the material anisotropy that mixes field components even in the bulk, but is further complicated by the boundary conditionsn \cite{Landau,Auld}. Some analytical solutions are only available for isotropic and non-piezoelectric wave guides \cite{SoundWaveguide}. For our purposes, however,  no explicit form of the eigen functions is needed but rather their general properties. 

Restricting for simplicity to optically isotropic materials, the optical eigen functions are found from the equation, 
\begin{eqnarray}
[\Delta_\perp - k^2] \psi_\al^n(r_\perp, k) = - \,(\epsilon\omega^2_n(k)/ c^2)\psi_\al^n(r_\perp, k) \,,
\end{eqnarray}
complemented with appropriate boundary conditions, $\omega_n(k)$ is the frequency of the $n$-th  mode. 

Equation for the acoustic eigenmodes is derived by eliminating piezoelectric potential using second equation in \Eq{EOMu} yielding,  
\begin{eqnarray}
 \hat {\cal D}_{\al\ga}[q]\phi_\ga ^m(r_\perp, q) = - 
 \rho \, \Omega^2_m (q)\phi_\al^m (r_\perp, q) \,,
\end{eqnarray}
where 
\begin{eqnarray}
\!\hat {\cal D}_{\al\ga}[q] = \!\left[c_{\al\be\ga\de} - {4\pi e_{\lambda\mu\al}
e_{\be\ga\de}\over \epsilon} \partial_{\lambda}\partial_\mu(\Delta_\perp - q^2)^{-1}\right]
\partial_{\be}\partial_\de ,  \nonumber
\end{eqnarray}
is a non-local operator with $\partial_x = iq$, and $\Omega_m (q)$ is the frequency of the $m$-th transverse mode. The solutions are generally complex and obey the symmetry relation,
$\phi_\al^m (r_\perp, -q) = [\phi_\al^{m} (r_\perp, q)]^{\ast}$. 

Within the eigen mode representation, the free field part of Lagrangian in \Eq{Lsimple} has the form, 
\begin{eqnarray}\label{L0mode}
{\cal L}_0 &=& {\epsilon \over 8\pi c^2} \sum_{k,n} \left\{  \dot A_n(k)  \dot A_n(-k)
- \omega^2_n(k)  A_n(k) A_n(-k) \right\} \nonumber\\
&+& {\rho\over 2} \sum_{n,q} \left\{  \dot u_n(q)  \dot u_n(-q) -
 \Omega^2_n(q)  u_n(q) u_n(-q) \right\}, 
\end{eqnarray}
while the Lagrangian of the photo-elastic interaction reads, 
\begin{eqnarray}\label{Eint} 
&&{\cal L}_{int} =  {1\over 8\pi c^2}\sum_{\substack{nn'm\\kk'q}} {\cal M}_{nn'm}^{kk'q} \,
{ \dot A_n(k) \dot A_{n'} (k') } u_m(q),   \\
&&{\cal M}_{nn'm}^{kk'q} 
=  \ga_{\al\be\ga\de}\,  \int {dV\over \sqrt{L^3}}\,
\psi_\al^n(r_\perp,k) \psi_\be^{n'}(r_\perp,k')\,\nonumber\\
&& \quad \quad \quad\quad \times \,e^{i(k+k')x} 
\left(\phi_\ga^m(r_\perp,q) e^{iqx}\right)_{,\de} \label{M} \,.
\end{eqnarray}
%

The major contribution to the interaction is given by the resonant wave triads selected by the resonance conditions, $\omega_{n'}(k')- \omega_{n}(k) = \Omega_m(q)$ and $ k'-  k =  q$. These equations define the scattering geometry with the optical modes propagating in opposite directions and having wave vectors, 
$k' \approx -k = (q/2)(1 + {\cal O}(\Omega/\omega))$, $\Omega \ll \omega$; the acoustic mode propagates along the direction of optical mode with larger frequency. Truncated to the resonant subspace, the Lagrangian (\ref{Eint}) reduces to the sum over resonant triads, each contribution 
having the form of the resonant three-wave interaction, 
\begin{eqnarray}\label{Hsbs}
\!\!\!\!\!{\cal L}_{int} \sim  A_{n}(k) A_{n'}^\ast (k') u_{m}(q) + A_{n}^\ast (k) A_{n'}(k') u_{m}^\ast (q) .
\end{eqnarray}
This interaction is generally known in theory of nonlinear waves \cite{Ablowitz}, and it
 describes scattering between two optical modes with emission and absorption  of acoustic mode. We note that the frequencies of the optical modes in \Eq{Hsbs} obey inequality, $\omega_{n'k'}
> \omega_{nk}$.


\subsection{SBS in quantum regime}

To proceed with the quantum description of SBS, we consider one of the optical modes to have much larger amplitude than the other and treat this pumping mode as a classical coherent state characterized by the number $N_p$ of pumping photons, $A_p \propto \sqrt N_p e^{-i\omega_p t}$. Then we apply the quantization procedure to weak optical and acoustic modes.

The quantization  is performed, first, by deriving  the Hamiltonian in the mode representation using \Eqs{L0mode} and (\ref{Eint}); then complex dimensionless quadratures are introduced via canonical 
transformation, 
\begin{eqnarray}
A_n(k) &=&  \sqrt { 2\pi \hbar c^2 / \epsilon\omega_n(k)}\,(a_{nk} + a_{n (-k)}^\ast) \,, \nonumber\\ 
u_m(q) &=& \sqrt{\hbar /2\rho\Omega_m(q)}(b_{mq} + b^\ast_{m(-q)}) \,,
\end{eqnarray}
and the canonical commutation relations are imposed on these quadratures, $[a_{nk}\,, a_{nk}^\dag]=1$,  and $[b_{mq}\,, b_{mq}^\dag]=1$. 


Selection of the pumping mode leads to the two different types of quantum Hamiltonians depending on whether frequency of the pumping mode is smaller or larger than the frequency of the optical signal mode.  In the first case, $\omega_p = \omega_s - \Omega$ (pumping mode is $A_{n}(k)$ in \Eq{Hsbs}),  the scattering  occurs with absorption of phonon into a blue shifted (anti-Stokes) sideband. This process results in the coherent phonon-photon conversion and is described with the beam splitter type Hamiltonian, 
\begin{eqnarray}\label{Hconversion}
{\cal H}_{SBS} =  - \hbar \sqrt N_p \left(  g_0 e^{-i\omega_p t} \,a^\dag b +  g_0^\ast e^{i\omega_p t} 
\,a b^\dag \right)\,. 
\end{eqnarray}
In the second case, $\omega_p = \omega_s + \Omega$ (mode $A_{n'}(k')$  is chosen as the pump),
 the scattering occurs with 
emission of phonon to a red shifted (Stokes) sideband. In this case, the SBS Hamiltonian takes the form of parametric amplifier,
\begin{eqnarray}\label{Hamplification}
{\cal H}_{SBS} =  - \hbar \sqrt N_p \left( g_0 e^{-i\omega_p t} \,a^\dag b^\dag +  g_0^\ast e^{-i\omega_p t} \,ab   \right)\,, 
\end{eqnarray}
and describes the amplification and two-mode squeezing of the optical and acoustic modes  \cite{Milburn}.

In both cases, the acousto-optic coupling  is  given by the vacuum phonon-photon coupling rate, $g_0$, %
\begin{eqnarray}\label{g0M}
 &&  g_0 =  {\cal M}  \sqrt{\hbar\omega_p\omega_s\Omega\over 32 \epsilon^2\rho s^2}, 
 \end{eqnarray}
enhanced by the pumping field. 
Here ${\cal M}$ is the  overlap integral in \Eq{M}   truncated to the resonant subspace,
\begin{eqnarray}
&& {\cal M} =   {\ga_{\al\be\ga\de}\over \sqrt L}  \times  \\
 && \int dV_\perp \psi_\al^p(r_\perp,q/2) \psi_\be  (r_\perp, q/2)
(\partial_\de/q + i\delta_{\delta x})\phi_\ga (r_\perp,q)\,,  \;\nonumber
\end{eqnarray}
superscript $p$ indicates the pumping mode. For an optimal design of the resonator, ${\cal M}$ is estimated, 
\begin{eqnarray}\label{Mest}
{\cal M}\sim \gamma / \sqrt V,
\end{eqnarray}
where $\gamma$ is a representative value of the photo-elastic tensor, and $V$ is the resonator 
volume.

For piezoelectric materials with relatively large photo-elastic interaction, such as LiNbO$_3$ and GaAs, the vacuum coupling rate  is estimated, $g_0  \sim $ 1.7 MHz$/\sqrt{V[\mu^3]}$ (LiNbO$_3$), and $g_0 > $ 6 MHz$/\sqrt{V[\mu^3]}$ (GaAs), while for AlN it is below 100 KHz due to small photo-elastic constant.

Our estimate for the vacuum coupling rate is solely based on the consideration of the photo-elastic interaction. In literature an additional mechanism of acousto-optic coupling is considered stemming from the displacement of resonator boundaries \cite{boundaryPainter,PainterBook,Chan}. Although this effect is described with a similar non-linear three-wave interaction to the one in \Eq{Hsbs} \cite{Law,ClerkBook}, the underlying physics is different in both cases: The photo-elastic effect affects the light velocity, while the boundary displacement changes geometric quantization of the cavity modes. Furthermore the  boundary effect in its generic form \cite{Law,boundaryHill} assumes a sharp step-wise boundary whose displacement results in a large change of dielectric constant that cannot be described with the photo-elastic effect. 
However, in practice the boundaries of integrated solid state wave guides are  smooth on the scale of the zero point acoustic displacement hence variation of dielectric constant under elastic deformation is small.     This justifies the photo-elastic approximation not only in the bulk but also at the boundary, while the optomechanical boundary effect  does not play a separate role. 

\subsection{Conversion efficiency}

 To evaluate the fidelity and efficiency of the phonon-photon conversion, we assume the qubit well detuned from the acoustic resonance, and consider the transducer as a 4-port device having optical and microwave input and output ports. 
 For the optical ports, the input-output relation   has conventional form \cite{Gardiner},
 $a_{out} = a_{in} - i \sqrt{2\kappa_0} \,a$, where $\kappa_0$ is a coupling rate  to the optical fiber. 
Considering the microwave ports,  we introduce the microwave field operators in the external transmission line, $c_{in}$ and $c_{out}$, and write the input-output relation taking advantage of the linear piezoelectric coupling of microwave and acoustic fields,  $c_{out} = c_{in} - i \sqrt{2\Gamma_0} \,b$, where $\Gamma_0$ includes the coupling rates of the IDT and the capacitive 
connection, $C_c$, to the transmission line, Fig. 1. 

The intra cavity field operators  satisfy the Langevin equations, associated with Hamiltonian (\ref{Hconversion}), which have the form in the interaction representation,  
\begin{eqnarray}\label{Langevin}
i\dot a + (\delta\omega  + i\kappa) a + g_0\sqrt{N_p} b &=& \sqrt{2\kappa_0}\, a_{in}\nonumber\\
i\dot b + (\delta\Omega  + i\Gamma) b + g^\ast_0\sqrt{N_p} a &=& \sqrt{2\Gamma_0} \,c_{in}\,.
\end{eqnarray}
Here we introduced detunings of the input optical signal, $\delta\omega$, and acoustic signal, $\delta\Omega$,  from the respective resonances, that satisfy relation, $\delta\omega - \delta\Omega =2\delta$, with $2\delta$ referring to detuning of the pump, $\omega_p=\omega_s  - \Omega + 2\delta$; $\kappa$ and 
$\Gamma$ denote total optical and acoustic damping rates, respectively. 

\begin{figure}[h]
\begin{center}
\includegraphics[width=0.3\textwidth]{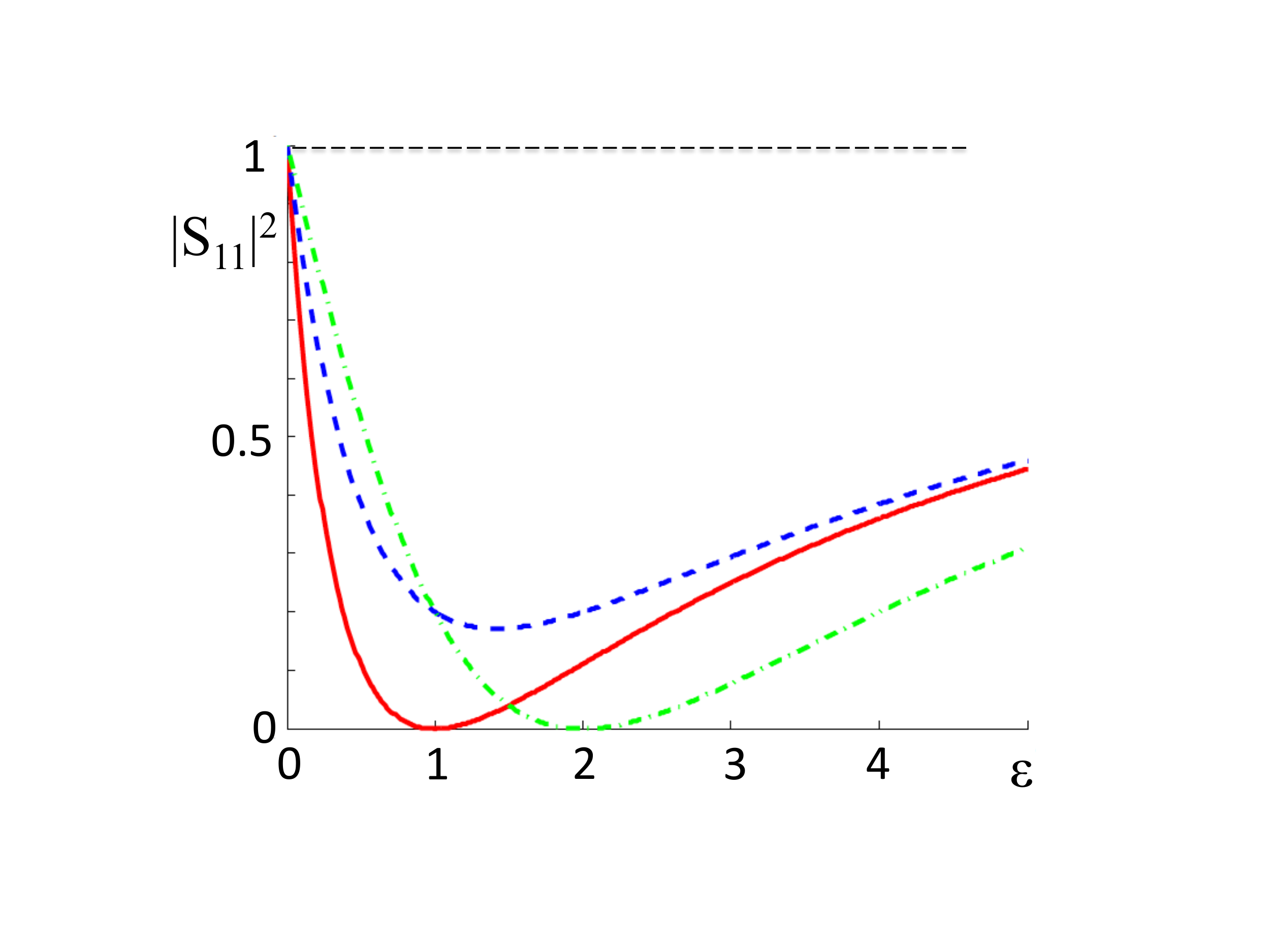}
\caption{(Color on-line) Conversion efficiency as function of normalized pumping strength, $\varepsilon = |g_0|^2N_p / \kappa_0\Gamma_0$, for different pump and signal detunings, internal losses are neglected.    Solid (red) line:  exact 
resonance, $\delta=\delta\omega_s=\delta\Omega=0$; dashed-dotted (green) line:  
$\delta\omega_s = \kappa_0$, $\delta\Omega = \Gamma_0$, $\delta = (\kappa_0 - \Gamma_0)/2$; dashed (blue) line: $\delta =0$, $\delta\Omega = \delta\omega_s = \Gamma_0$. 
}
\label{}
\end{center}
\end{figure}
Solving these equations and substituting into the input-output relations, we compute the scattering matrix and evaluate the transmission and reflection amplitudes, 
\begin{eqnarray}\label{S}
  S_{11} &=&  1- 2i\kappa_0 ( \delta\Omega +i\Gamma ) /  D \nonumber\\ 
  S_{22} &=&  1- 2i\Gamma_0 ( \delta\omega +i\kappa ) /  D  \\
  S_{12} &=&  -2i g_0\sqrt{N_p\kappa_0\Gamma_0}/  D \nonumber, 
\end{eqnarray}
$ D = (\delta\omega +i\kappa)(\delta\Omega + i\Gamma) - |g_0|^2N_p$. 

In the limit of negligibly small internal losses, $\kappa - \kappa_0 \ll \kappa_0 $, and 
$\Gamma - \Gamma_0 \ll \Gamma_0$, the scattering matrix acquires a unitary form with 
$|S_{11}| = |S_{22}| $, and 
$|S_{12}|= |S_{21}| = 1 - |S_{11}|^2$, 
thus providing reversibility of phonon-photon conversion. 

Furthermore, a full conversion indicated by  absence of the  reflection,   
 $|S_{11}| =|S_{22}| =0$, is achieved for 
\begin{eqnarray}\label{g>}
 |g_0|^2 N_{p}\geq \kappa_0\Gamma_0 \,,
\end{eqnarray}
the equality occurring at the exact resonance, $\delta\omega=\delta\Omega = 0$, as illustrated in Fig. 2. This equation defines the minimum pump strength required for the full phonon-photon conversion. For realistic Q-factor values, $Q_{opt}=\omega_s/\kappa \sim 10^5$ and $Q_{ac}= \Omega/\Gamma \sim 10^4$, the corresponding pump photon density is estimated, $N_p/V \sim 10^4$ photons/$\mu^3$, for LiNbO$_3$. It is interesting, that this minimum value coincides with  the threshold of parametric oscillation in the Stokes scattering channel, where the amplified phonon-photon vacuum noise reaches its maximum value \cite{Waltraut,Simoen}.
Above the threshold, the full conversion can be achieved for finite detunings, 
$\delta\omega= \kappa_0(|g|^2 / \kappa_0\Gamma_0 -1)^{1/2}$, and $\delta\Omega = (\Gamma_0/\kappa_0)\delta\omega$.

\begin{figure}[h]
\begin{center}
\includegraphics[width=0.3\textwidth]{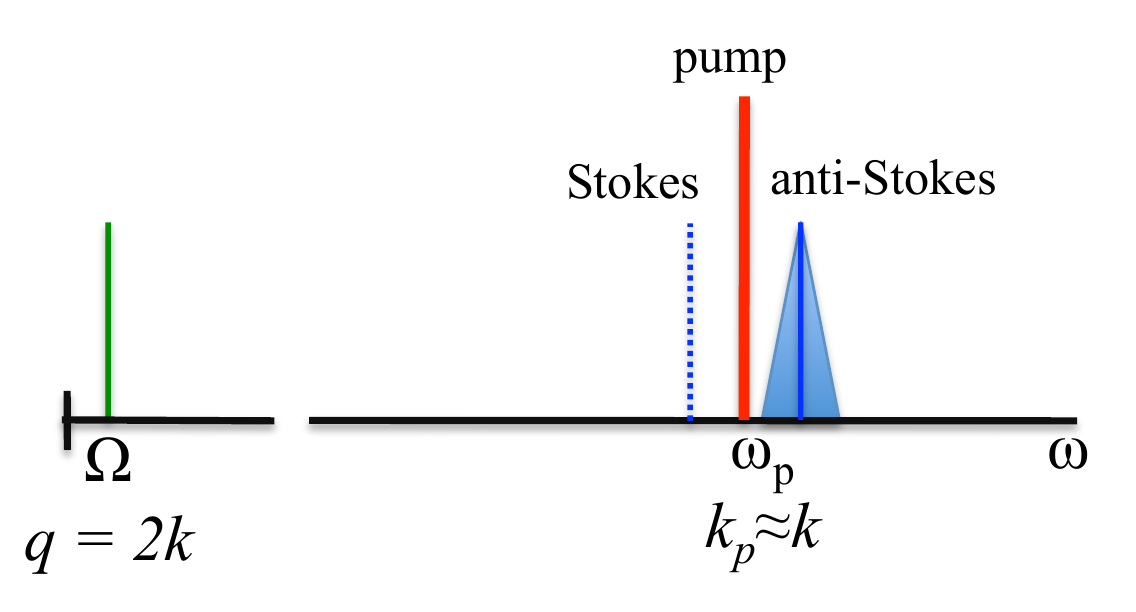}
\caption{(Color online) Frequency diagram of the SBS: the scattering occurs into two sidebands symmetrically shifted from the pump frequency by $\pm\Omega$, blue triangle indicates broadening of the cavity resonance,   $\kappa_0<2\Omega$.  }\label{}
\end{center}
\end{figure}
%

For the purpose of faithful single quantum phonon-photon conversion, the Stokes scattering is 
an undesirable process, since it generates spurious phonon-photon pairs out of the vacuum fluctuations. 
The Stokes scattering can be suppressed by making Stokes sideband frequency well detuned from the 
cavity resonance. This imposes a constraint on the resonance width, $\kappa < 2 \Omega$, 
corresponding to the resolved side band regime, as illustrated in Fig. 3. Accordingly  the optical 
Q-factor has a lower bound,   $Q_{opt} > \omega_s/2\Omega \sim 2\cdot 10^4$. 
Additional possibility for suppressing the Stokes scattering exists in the ring resonator of Fig. 1b: taking advantage of the asymmetry  of the Brillouin scattering, one may implement asymmetric IDT to selectively emit and absorb acoustic waves moving in the direction only supporting the anti-Stokes 
scattering.

\section{Practical implementation}

The transducer performance   - fidelity and efficiency of the phonon-photon conversion - depends on a number of material and device parameters. The most important limitation on these parameters is imposed  by the requirement of minimum heating effect of the pump losses, which is damaging for a fragile cryogenic environment of c-QED devices. This implies minimization of the pump power, \Eq{g>}, required for the full conversion. This is achieved by reducing external optical and acoustic losses, and 
choosing materials with large phonon-photon vacuum coupling rate. However, the engineered external losses cannot be made arbitrary small - they must significantly exceed the internal losses in order to maintain the unitarity of the conversion,  \Eq{S}.  For the optical internal damping at the level of $\sim 0.1\,\kappa_0 \sim$ 1 GHz, the estimated dissipated power density within the resonator would be $\sim 1\, \mu$W/$\mu^3$ (for LiNbO$_3$).  

Furthermore, to maximally reduce the heating effect, it is desirable  to confine the losses to the interaction volume  within the resonator, while keeping low pump power in the feeding fiber. To achieve this one needs to engineer additional narrow resonance at the pump frequency, then estimated pump power would be within the range of 10 $\mu$W. The resonator with two tightly spaced and resolved optical resonances could be realized using method of coupled cavities \cite{Yariv}.

The value of the vacuum coupling rate, \Eq{g0M}, is essentially defined by the  photo-elastic coefficient, and the resonator volume. The  photo-elastic interaction is quantified in literature with the Pockel coefficient $p$, related to our coefficient, $\gamma= pn^4$,  where $n$ is the refractive index. For piezoelectric materials of interest, the largest Pockel coefficients vary from $p=0.02$, in AlN, to $p > 0.16$, in GaAs and LiNbO$_3$. Correspondingly, the variation of 
$\gamma$ is in the range 0.3 - 20. Since the minimum  pump power is proportional to the squared photo-elastic coefficient, the required pump power may differ by up to three orders of magnitude depending on the choice of material.

The cavity volume is to be reduced to maximize the vacuum coupling rate. In this respect, the optomechanical crystal resonators \cite{PainterBook,Cleland,Srinivasan} seem to provide an ultimate solution, having volume of few cubic wave lengths. The integrated ring resonator depicted in Fig. 1b, is more favorable compared to the planar SAW device in Fig. 1a since it may have small transverse dimensions comparable to the wave length, while the width of the planar SAW resonator is of the order of thens wave length (although this can be reduced by using focusing mirrors). Furthermore, planar resonator has a large length, up to thousand wave lengths \cite{Leek}, due to weak localization effect of metallic fingers of the Bragg mirrors that induce very small modulation of the SAW velocity \cite{Datta}.

\section{Conclusion}

We proposed and performed theoretical analysis of reversible quantum transducer for coupling  microwave and optical photons. The transducer employs acoustic phonons in GHz frequency range as an intermediate agent, and consists of an integrated acousto-optic cavity fabricated with piezoelectric material. The phonon-optical photon conversion is provided by the mechanism of stimulated Brillouin scattering (SBS),  while the phonon-microwave photon conversion is due to the piezoelectric effect. We find that the SBS induced vacuum coupling rate is in the range of few MHz per $\mu^3$ cavity volume, and the full phonon-photon conversion can be achieved at pump power of tens $\mu$W. Our analysis of the material and device parameters that would provide full and faithful quantum phonon-photon conversion are feasible for implementation with the state of the art integrated acousto-optics.


\section* {Acknowledgement} 
 The author is thankful for useful discussions to Magnus Karlsson, Martin Gustafsson, Per Delsing, Thomas Aref, Peter Leek, Victor Torres, and G\"oran Wendin. 
Support from the European FP-7 Consortium ScaleQIT is gratefully acknowledged.

\widetext
\appendix
\section{Derivation of classical SBS equations }
The starting point is the macroscopic action for acousto-optic medium consisting of electromagnetic and elastic parts,
\begin{eqnarray}\label{L}
S=\int dt\; {\cal L}, \quad
{\cal L} =  {1\over 2}\int dV \left({E^2 - H^2\over 4\pi}  + \varphi P_{\al,\al} + {1\over c} A_\al \dot P_\al   + \rho \dot u^2 - c_{\al\be\ga\de} u_{\al,\be} u_{\ga,\de}\right) \,, 
\end{eqnarray}
Macroscopic polarization, $P_\al = \chi_{\al\be}  E_\be - 2e_{\al\be\ga}u_{\be,\ga} +  
(1/ 4\pi)\ga_{\al\be\ga\de} E_\be u_{\ga,\be}$, contains piezoelectric and photo-elastic interaction terms, $E_\al = -\varphi_{,\al} - (1/c)\dot A_\al$, $ H_\al = ({\rm rot}\, A)_\al$, 
magnetic effects are not included in this calculation.
The tensor coefficients entering the Lagrangian possess symmetries,
\begin{equation}\label{sym}
e_{\al\be\ga}= e_{\al\ga\be}; \quad  \ga_{\al\be\ga\de} = \ga_{\ga\de\al\be} = \ga_{\be\al\ga\de} = 
\ga_{\al\be\de\ga}, \quad c_{\al\be\ga\de} = c_{\ga\de\al\be} = c_{\be\al\ga\de} = 
c_{\al\be\de\ga}\,.
\end{equation}
\\
Variation of the electromagnetic part of the Lagrangian is conveniently done in two steps. First we  compute variation over variable $E_\al$, that explicitly enters the Lagrangian, 
\begin{eqnarray}
\de{\cal L}_E = {1\over 4\pi} \int dV \,\left[ (\epsilon_{\al\be} - 2\pi\chi_{\al\be} + \ga_{\al\be\ga\de} u_{\ga,\de} ) E_\be \right] \de E_\al, \quad 
\epsilon_{\al\be} =  \de_{\al\be} + 4\pi\chi_{\al\be}\,.
\end{eqnarray}
Then we express this variation in the terms of variations of electromagnetic potentials,
\begin{eqnarray}
\de{\cal L}_E &=& {1\over 4\pi} \int dV \,\left\{ \left(\epsilon_{\al\be}\dot E_\be - 2\pi\chi_{\al\be}\dot E_\be + 
{1\over 2}\ga_{\al\be\ga\de}\partial_t ( u_{\ga,\de} E_\be) \right) {1\over c}\de A_\al \right.\nonumber\\
&+& \left. \left(\epsilon_{\al\be} E_{\be,\al} - 2\pi\chi_{\al\be} E_{\be,\al} 
 + {1\over 2}\ga_{\al\be\ga\de} ( u_{\ga,\de} E_\be)_{, \al} \right)\de \varphi \right\} \,. %
\end{eqnarray}
The next step is to compute the variation due to explicitly entering \Eq{L} variables $A_\al$ and $\varphi$, 
\begin{eqnarray}
\de{\cal L}_{A,\varphi} &=& {1\over 4\pi} \int dV\, \left\{ {1\over c} \left( 2\pi \chi_{\al\be} \dot E_\be - 
4\pi e_{\al\be\ga}\dot u_{\be,\ga} +  
{1\over 2}\ga_{\al\be\ga\de} \, \partial_t(E_\be u_{\ga,\de})   - c({\rm rot} H)_\al \right) \de A_\al  \right. \nonumber\\
&+&\left.  \left( 2\pi\chi_{\al\be} E_{\be,\al} - 4\pi e_{\al\be\ga} u_{\be,\ga\al} +  
{1\over 2}\ga_{\al\be\ga\de} (E_\be u_{\ga,\de})\right)\!_{,\al} \,\de\varphi  \right\}\,. %
\end{eqnarray}
The total variation of the electromagnetic part of the Lagrangian consists of the sum of these two parts, $\delta {\cal L}_{Em} = \delta {\cal L}_{E} + \delta {\cal L}_{A,\varphi}$, 
\begin{eqnarray}\label{Lem}
\delta {\cal L}_{Em}  &=& {1\over 4\pi} \int dV\, \left\{\left( {1\over c}\dot D_{\al} - (\rt \,H)_\al \right)\de A_\al + D_{\al,\al}\de \varphi \right\} \,,
 \end{eqnarray}
conveniently written through the  electric displacement defined by equation, 
\begin{eqnarray}\label{Dapp}
D_\al &=& \epsilon_{\al\be} E_{\be} - 4\pi e_{\al\be\ga} u_{\be,\ga} +  
\ga_{\al\be\ga\de} \, E_\be u_{\ga,\be} \,.
\end{eqnarray}
\\
Variation of the Lagrangian over elastic displacement yields,
\begin{eqnarray}\label{Lel}
\delta {\cal L}_{El}  &=& \int dV\, \left\{- \rho \ddot u_{\ga} + c_{\al\be\ga\de}u_{\al,\be\de} +  e_{\al\be\ga} E_{\al,\be} -  {1\over 8\pi}\ga_{\al\be\ga\de}(E_{\al} E_{\be})_{,\de} \right\} \de u_\ga 
\,.
\end{eqnarray}
\\
From \Eqs{Lem} and (\ref{Lel}) we extract equations of motion,
\begin{eqnarray}\label{EOM1}
&&\dot D_{\al} = c\,(\rt \,H)_\al,  \quad c\,(\rt E)_\al = -\,\dot H_\al,  \quad D_{\al,\al}=0 \\
\label{EOM2} &&\rho \ddot u_{\ga} = c_{\al\be\ga\de}u_{\al,\be\de} +  e_{\al\be\ga} E_{\al,\be} -  {1\over 8\pi}\ga_{\al\be\ga\de}(E_{\al} E_{\be})_{,\de} \,,%
\end{eqnarray}
which are presented in the main text, \Eqs{EOME}-(\ref{D}).%

Equations of motion can be significantly simplified due to the fact that the optical field has much larger phase velocity than the acoustic field. For comparable wave vectors of both fields this implies that the time variation of the transverse optical field, $A_\al$, is much faster than the time variation of the acoustic field, $u_\al$, and the related longitudinal piezoelectric field, $\varphi$. Therefore one may omit small terms in \Eq{EOM1} proportional to the time derivatives of $u_\al$, and $\varphi$, giving
\begin{eqnarray}\label{EOM1simple}
\epsilon_{\al\be}\ddot  A_{\be} -{c^2\over \mu} \Delta \,A_\al +  \ga_{\al\be\ga\de} \, \ddot A_\be u_{\ga,\de} =0 \,.
\end{eqnarray}
Averaging over fast temporal oscillation in \Eq{EOM2} and \Eq{Dapp} eliminates the linear terms proportional to $A_\al$, but retains the quadratic term essential for the SBS, 
\begin{eqnarray}\label{EOM2simple}
 &&\rho \ddot u_{\ga} - c_{\al\be\ga\de}u_{\al,\be\de} +  e_{\al\be\ga} \varphi_{,\al\be} +  {1\over 8\pi c^2}\ga_{\al\be\ga\de}(\overline{\dot A_{\al} \dot A_{\be}})_{,\de}
 + {1\over 8\pi }\ga_{\al\be\ga\de}(\varphi_{,\al} \varphi_{,\be})_{,\de}=0 \nonumber\\%
 && -\epsilon_{\al\be} \varphi_{,\be\al} - 4\pi e_{\al\be\ga} u_{\be,\ga\al} -
 \ga_{\al\be\ga\de}(\varphi_{,\be} u_{\ga,\de})_{,\al} = 0  
\end{eqnarray}
Finally, bearing in mind that under SBS only the optical field contains large amplitude component, while 
piezo-elastic fields are weak, we omit the last, nonlinear terms in the both lines of \Eq{EOM2simple}.
The resulting equations are presented in the main text, \Eqs{EOMA}-(\ref{EOMu}).


 
\end{document}